\documentclass[runningheads]{llncs}
\usepackage[T1]{fontenc}
\usepackage{cite}
\usepackage{tikz}
\usepackage{amsmath,bm}
\usepackage{algorithm}
\usepackage{algpseudocode}
\PassOptionsToPackage{hyphens}{url}\usepackage{hyperref}
\usepackage{amssymb}
\usepackage{float}
\usepackage{listings}
\usepackage[T1]{fontenc}
\definecolor{myblack}{RGB}{0, 0, 0}
\definecolor{mygreen}{RGB}{0, 146, 146}
\definecolor{myblue}{RGB}{0, 110, 219}
\definecolor{myred}{RGB}{146, 0, 0}

\definecolor{darkred}{RGB}{193, 39, 45}
\definecolor{indigo}{RGB}{0, 0, 167}
\definecolor{teal}{RGB}{0, 129, 118}
\definecolor{yellow}{RGB}{238, 204, 22}
\definecolor{lightgray}{RGB}{179, 179, 179}

\colorlet{potrf}{darkred}
\colorlet{trsm}{teal}
\definecolor{syrk}{RGB}{238, 204, 22}
\colorlet{gemm}{indigo}
\definecolor{utile}{rgb}{0.8, 0.8, 0.8}

\lstdefinestyle{inlinecpp}{
  language=C++,
  basicstyle=\ttfamily,
  keywordstyle=\bfseries,
  commentstyle=\slshape,
  stringstyle=\ttfamily,
  identifierstyle=,
}

\lstdefinestyle{inlinesh}{
  language=sh,
  basicstyle=\ttfamily,
}
\usepackage{graphicx}
\usepackage{enumitem}
\setlist[itemize]{label=\textbullet}

\begin{document}
\title{From Fork-Join to Asynchronous Tasks: Parallelizing Tiled Cholesky Decomposition with OpenMP and HPX}
\titlerunning{From Fork-Join to Asynchronous Tasks}
%
\author{Alexander Strack\inst{1}\orcidID{0000-0002-9939-9044} \and
Alexander Van Craen\inst{1}\orcidID{0000-0002-3336-7226} \and
Dirk Pfl\"uger\inst{1}\orcidID{0000-0002-4360-0212}}
%
\authorrunning{A. Strack et al.}

\institute{Institute of Parallel and Distributed Systems, University of Stuttgart,\\ 70569 Stuttgart, Germany\\
\email{\{alexander.strack, alexander.van-craen, dirk.pflueger\}@ipvs.uni-stuttgart.de}}

\maketitle              
\begin{abstract}
	Fork-join parallelism, popularized by OpenMP, remains the dominant model for shared-memory parallel programming, but its implicit synchronization barriers can penalize algorithms with inhomogeneous workloads.
	Asynchronous many-task (AMT) runtimes sidestep these barriers by expressing work as a dependency graph of fine-grained tasks. Yet, the actual performance benefit over a carefully written fork-join baseline is rarely quantified.
	In this work, we introduce Cholesky-Bench and use it to revisit the tiled Cholesky decomposition, a canonical irregular kernel, comparing four parallelization variants of the right-looking algorithm across two runtimes: the OpenMP implementations shipped with GCC and LLVM, and the HPX AMT runtime.
	The variants span classical fork-join, a collapsed fork-join that exposes additional inner-loop parallelism, synchronous tasking, and asynchronous tasking with explicit data dependencies.
	We benchmark all eight combinations on a dual-socket 128-core AMD Zen 2 node across multiple tile sizes and problem sizes.
	Our results show that across all variants, HPX outperforms OpenMP at the optimal tile size by 15\%--30\%.
	Specifically, asynchronous HPX tasks are up to 26\% faster than their OpenMP counterparts, and exhibit roughly $3.8\times$ smaller task overhead.
	Furthermore, the collapsed fork-join variants close most of the gap to synchronous tasking.
	Removing redundant synchronization barriers yields an additional improvement of 7\% (OpenMP) to 14\% (HPX).
	A GCC-versus-LLVM comparison further reveals compiler-specific differences in fork-join scheduling and task-creation overheads.

\end{abstract}
\section{Introduction}

Fork-join parallelism became the de facto shared-memory programming model with the introduction of the OpenMP directive-based standard~\cite{Dagum1998_openmp}.
While it is highly effective for homogeneous, embarrassingly parallel workloads, more complex numerical algorithms frequently exhibit inhomogeneous work distributions.
Because fork-join relies, by design, on implicit synchronization barriers at the end of each parallel region, it can leave compute resources idle whenever the work inside the region is unbalanced.

To address this limitation, an alternative parallelization paradigm emerged, in which the overall computation is decomposed into fine-grained tasks that can be dynamically scheduled onto available resources.
When augmented with explicit task dependencies, this model removes redundant synchronization barriers and lets the runtime execute independent work asynchronously, typically improving both load balance and resource utilization.

A canonical problem in this setting is the solution of symmetric positive-definite linear systems via the Cholesky decomposition. The decomposition arises across a wide range of domains, including geostatistics~\cite{Krige1951_kriging, Abdulah2018_exageostat_starpu_cholesky}, machine learning~\cite{Rasmussen2005_gp_ml, Kocijan2016_gp_si_book, Helmann2026_gprat}, and scientific computing~\cite{Abdulah2024_parsec_cholesky_climate_gpu}, and its triangular dependency structure exposes abundant fine-grained parallelism with inherently irregular workloads, making it a natural benchmark for task-based runtimes.

Despite the problem's popularity, the practical performance gains of asynchronous tasking over synchronous tasking or a carefully tuned fork-join baseline are rarely quantified side-by-side in the literature.
Thus, in this work, we present Cholesky-Bench, an extensible framework for comparing parallelization approaches across several libraries.
Cholesky-Bench includes multiple fork-join and tasking-based parallelization variants of the tiled Cholesky decomposition implemented both on top of the OpenMP standard~\cite{Dagum1998_openmp} and on top of the asynchronous many-task runtime HPX~\cite{Kaiser2020_hpx}.
Our specific contributions are:
\begin{itemize}
	\item Cholesky-Bench, an open framework for comparing parallelization strategies for the tiled Cholesky decomposition under a unified software stack;
	\item a head-to-head comparison of fork-join, synchronous, and asynchronous tasking across a wide range of tile sizes for both OpenMP and HPX;
	\item a problem-size scaling study at fixed tile counts that isolates the task-management overheads of each runtime; and
	\item a comparison of the OpenMP libraries shipped with the GCC and LLVM compilers on the same source code.
\end{itemize}

The remainder of this paper is organized as follows.
Section~\ref{sec:related_work} surveys related work on Cholesky decomposition with OpenMP and with asynchronous many-task runtimes.
Section~\ref{sec:methods} introduces the tiled Cholesky algorithm, the four parallelization variants, and their implementations in both OpenMP and HPX.
Section~\ref{sec:results} presents our benchmark results, and Section~\ref{sec:conclusion} concludes with an outlook on future work.

\section{Related Work}\label{sec:related_work}

OpenMP and HPX are by no means the only runtimes providing support for asynchronous tasking.
Beyond the task primitives included in the C\texttt{++} Standard Library~\cite{Diehl2023_hpx_book}, several purpose-built asynchronous many-task (AMT) runtimes have been developed~\cite{Thoman2018_amt_survey}.
A recent survey by Schuchart et al.~\cite{Schuchart2025} discusses their design differences and surveys representative applications.
One of the few means of comparing performance and overhead across AMTs is Task Bench~\cite{Slaughter2020_taskbench}, which has been extended to additional runtimes in~\cite{Wu2023_hpx_taskbench, Lahnor2026_taskbench_itoyori_hpx}.
Several of the AMTs covered in this survey provide some form of tiled Cholesky implementation.

The PaRSEC runtime~\cite{Bosilca2013_parsec}, for example, has been used for a tiled Cholesky factorization that underpins a range of applications~\cite{Cao2022_parsec_cholesky_gp_cpu, Abdulah2024_parsec_cholesky_climate_gpu}.
Building on PaRSEC, the TTG framework~\cite{Bosilca2020_ttg} offers a template-based expression of task dependencies, which has been used to benchmark tiled Cholesky in~\cite{Schuchart_ttg_cholesky}.
The Chameleon library~\cite{Agullo2017_chameleon_cholesky}, built on top of the StarPU runtime~\cite{Augonnet2011_star_pu}, provides a tiled Cholesky decomposition that is employed, for instance, in ExaGeoStat~\cite{Abdulah2018_exageostat_starpu_cholesky}.
Dorris et al.~\cite{Dorris2016_cholesky_plasma_knl} compare three variants of the tiled Cholesky decomposition parallelized with asynchronous OpenMP tasking as part of the PLASMA library~\cite{Dongarra2019_plasma_openmp}.

For HPX, a tiled Cholesky decomposition was first developed in~\cite{Strack2023_cholesky}.
The GPRat library~\cite{Helmann2026_gprat} for Gaussian process regression builds on this work and is designed as an alternative to GPflow~\cite{Matthews2017_gpflow} and GPyTorch~\cite{Gardner2018_gpytorch}.
GPRat is portable across a wide range of hardware, from accelerators~\cite{Moellmann2026_gprat_cuda} to emerging architectures such as RISC-V~\cite{Strack2026_gprat_riscv}.
Further notable HPX applications include the astrophysics code Octo-Tiger~\cite{Marcello2021_octotiger}, the distributed FFT tool HPX-FFT~\cite{Strack2024_hpxfft}, and the radiation-hydrodynamics code HARD~\cite{Loiseau2025_hard}, which can use MPI~\cite{Mpi2012}, Legion~\cite{Bauer2012_legion}, or HPX as alternative backends through FleCSI~\cite{Bergen_flecsi_2, Strack2026_flecsi_hard, Herring2025_flecsi_hpx}.

As individual Cholesky implementations are typically embedded within libraries, applications, or example code, conducting a fair and reproducible comparison is cumbersome or even infeasible.
Cholesky-Bench addresses this gap by providing a unified and fair benchmarking baseline.

\section{Methods}\label{sec:methods}

In this section, we briefly introduce the tiled Cholesky decomposition and then present the different parallelization approaches based on fork-join and tasking.
We subsequently discuss their concrete implementations in OpenMP and HPX.

\subsection{Tiled Cholesky Decomposition}

The Cholesky factorization admits several parallelization strategies.
While a naive approach parallelizes the element-wise loops, more performant implementations rely on a block- or tile-based formulation~\cite{Buttari2009_parallel_exakt_solvers}.

The symmetric positive-definite matrix $\mathbf{A}$ is partitioned into square tiles, where $\mathbf{A}_{I, J}$ denotes the tile in block row $I$ and block column $J$.
Owing to symmetry, only the diagonal and the strictly lower triangular part of $\mathbf{A}$ need to be stored. The tiled Cholesky algorithm then operates on these tiles.
Three variants of the algorithm are commonly distinguished: right-looking, left-looking, and top-looking, where the name refers to the portion of the matrix accessed during the trailing submatrix update~\cite{Dorris2016_cholesky_plasma_knl}.
In this work, we focus on the right-looking variant (Figure~\ref{fig:tiled-algo}).
The algorithm updates the tiles of $\mathbf{A}$ in-place using BLAS and LAPACK routines:
\begin{itemize}
	\item \textcolor{potrf}{\textbf{POTRF}}: the \emph{\textcolor{potrf}{PO}sitive-definite \textcolor{potrf}{TR}iangular \textcolor{potrf}{F}actorization} performs an in-place Cholesky decomposition $\mathbf{A}_{J,J} \leftarrow \mathrm{chol}(\mathbf{A}_{J,J})$ on the diagonal tile.
	\item \textcolor{trsm}{\textbf{TRSM}}: the \emph{\textcolor{trsm}{TR}iangular \textcolor{trsm}{S}olve with \textcolor{trsm}{M}ultiple right-hand sides} solves the triangular systems $\mathbf{A}_{I,J} \leftarrow \mathbf{A}_{I,J} \mathbf{A}_{J,J}^{-\mathsf{T}}$ on the current block column ($I > J$).
	\item \textcolor{syrk}{\textbf{SYRK}}: the \emph{\textcolor{syrk}{SY}mmetric \textcolor{syrk}{R}ank-\textcolor{syrk}{K} update} updates each diagonal tile of the trailing submatrix via $\mathbf{A}_{I,I} \leftarrow \mathbf{A}_{I,I} - \mathbf{A}_{I,J} \mathbf{A}_{I,J}^{\mathsf{T}}$ for $I > J$.
	\item \textcolor{gemm}{\textbf{GEMM}}: the \emph{\textcolor{gemm}{GE}neral \textcolor{gemm}{M}atrix-\textcolor{gemm}{M}atrix multiplication} updates each off-diagonal tile of the trailing submatrix via $\mathbf{A}_{I,K} \leftarrow \mathbf{A}_{I,K} - \mathbf{A}_{I,J} \mathbf{A}_{K,J}^{\mathsf{T}}$ for $J < K < I$.
\end{itemize}

Figure~\ref{fig:tiled-vis} illustrates which operations act on which tiles for a $4\times 4$ tiled matrix during the first block-column iteration.

\begin{figure}[t]
	\centering
	\begin{minipage}[c]{0.49\textwidth}
		\begin{algorithmic}[1]
			\footnotesize
			\For{$J \gets 0$ \textbf{to} $M - 1$}
			\State \textcolor{potrf}{\textbf{POTRF}}($\mathbf A_{J,J}$)
			\For{$I \gets J + 1$ \textbf{to} $M - 1$}
			\State \textcolor{trsm}{\textbf{TRSM}}($\mathbf A_{J,J}, \mathbf A_{I,J}$)
			\EndFor
			\For{$I \gets J + 1$ \textbf{to} $M - 1$}
			\State \textcolor{syrk}{\textbf{SYRK}}($\mathbf A_{I,J}, \mathbf A_{I,I}$)
			\For{$K \gets J + 1$ \textbf{to} $I - 1$}
			\State \textcolor{gemm}{\textbf{GEMM}}($\mathbf A_{I,J}, \mathbf A_{K,J}, \mathbf A_{I,K}$)
			\EndFor
			\EndFor
			\EndFor
		\end{algorithmic}
		\caption{Pseudocode for right-looking tiled Cholesky decomposition of $\mathbf A$.}
		\label{fig:tiled-algo}
	\end{minipage}
	\hfill
	\begin{minipage}[c]{0.48\textwidth}
		\centering
		\begin{tikzpicture}[scale=0.9]
			\tikzstyle{every node}=[rounded corners=3pt, font=\large, minimum size=0.8325cm, inner sep=0pt]

			\draw (0,0) rectangle (4,-4); 
			\draw[step=1cm, densely dotted] (0,0) grid (4,-4);

			\node[draw=black, shape=rectangle, fill=darkred, line width=0.5] at (0.5, -0.5) {\textbf{P}};

			\node[draw=black, shape=rectangle, fill=teal, line width=0.5] at (0.5, -1.5) {\textbf{T}};
			\node[draw=black, shape=rectangle, fill=teal, line width=0.5] at (0.5, -2.5) {\textbf{T}};
			\node[draw=black, shape=rectangle, fill=teal, line width=0.5] at (0.5, -3.5) {\textbf{T}};

			\node[draw=black, shape=rectangle, fill=yellow, line width=0.5] at (1.5, -1.5) {\textbf{S}};
			\node[draw=black, shape=rectangle, fill=yellow, line width=0.5] at (2.5, -2.5) {\textbf{S}};
			\node[draw=black, shape=rectangle, fill=yellow, line width=0.5] at (3.5, -3.5) {\textbf{S}};

			\node[draw=black, shape=rectangle, fill=indigo, line width=0.5] at (1.5, -2.5) {\textcolor{white}{\textbf{G}}};
			\node[draw=black, shape=rectangle, fill=indigo, line width=0.5] at (1.5, -3.5) {\textcolor{white}{\textbf{G}}};
			\node[draw=black, shape=rectangle, fill=indigo, line width=0.5] at (2.5, -3.5) {\textcolor{white}{\textbf{G}}};
		\end{tikzpicture}

		\caption{Example for matrix $\mathbf A$ split into $4 \times 4$ tiles, colored by tasks in the first iteration ($J=0$): \textcolor{potrf}{\textbf{POTRF}}, \textcolor{trsm}{\textbf{TRSM}}, \textcolor{syrk}{\textbf{SYRK}}, and \textcolor{gemm}{\textbf{GEMM}}.}
		\label{fig:tiled-vis}
	\end{minipage}
\end{figure}

\subsection{Implementations}

\begin{figure}[t]
	\centering
	\includegraphics[width=\textwidth]{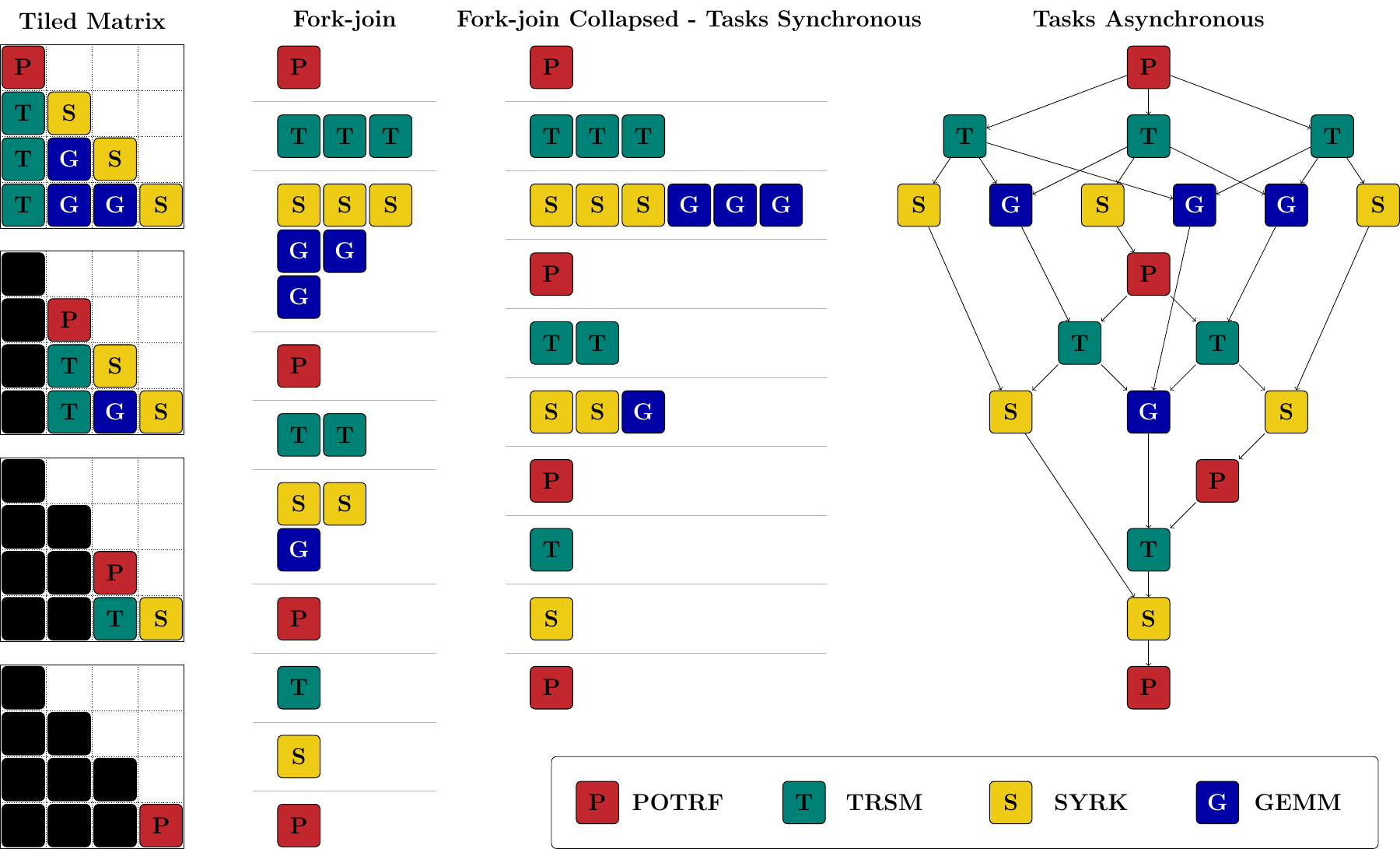}
	\caption{Comparison of different parallelization approaches for the tiled right-looking Cholesky decomposition on a $4 \times 4$ tiled matrix.
		On the left, the respective tasks are visualized within the tiled matrix.
		Grey horizontal lines denote implicit synchronization barriers imposed by the parallelization model, and arrows encode inter-task data dependencies.}
	\label{fig:variants}
\end{figure}

In the following, we present different approaches for parallelizing the tiled Cholesky algorithm (Figure~\ref{fig:tiled-algo}) using OpenMP and HPX, covering both fork-join and task-based parallelism.

\paragraph{OpenMP variants.}
The simplest way to parallelize the pseudocode in Figure~\ref{fig:tiled-algo} with OpenMP is to annotate the outer loops with \lstinline[style=inlinecpp]|#pragma omp parallel for|.
As OpenMP is integrated into major compilers such as GCC and LLVM, enabling parallel execution typically requires no more than adding the \texttt{-fopenmp} flag.
A drawback of this naive approach is that the amount of parallel work in the trailing submatrix update is limited solely by the outer loop.
When the outer loop contains only a few iterations, i.e., near the end of the factorization, and the inner loop is not exposed to the scheduler, threads are underutilized.
Dynamic scheduling of the outer loop alleviates this somewhat while there are enough iterations left, but cannot recover the parallelism hidden in the unexposed inner loop.

Alternatively, a remedy is to collapse the two submatrix-update loops using the \lstinline[style=inlinecpp]|collapse(2)| clause, exposing both dimensions of work simultaneously.
Because \lstinline[style=inlinecpp]|collapse| requires perfectly nested loops, we merge the \textcolor{syrk}{\textbf{SYRK}} and \textcolor{gemm}{\textbf{GEMM}} updates into a single loop nest and dispatch the correct routine via an \lstinline[style=inlinecpp]|if|-statement on the tile indices.
As Figure~\ref{fig:variants} illustrates, the collapsed fork-join variant exposes substantially more parallel work per phase than the naive version while retaining the same implicit synchronization barriers.

Starting with OpenMP 3.0\footnote{OpenMP specifications: \url{https://www.openmp.org/specifications/}, accessed 2026-06-10.}, the standard introduced basic task support: individual statements or scopes can be launched as asynchronous work units via \lstinline[style=inlinecpp]|#pragma omp task|.
Without further annotation, however, tasks execute independently of one another.
Correct program order must therefore be enforced with explicit \lstinline[style=inlinecpp]|#pragma omp taskwait| barriers.
In this model, the amount of exposed parallelism and the number of synchronization barriers are identical to those of the collapsed fork-join variant, so comparable performance is expected in principle.
Any difference between the two isolates the task-creation and scheduling overheads relative to fork-join.

In OpenMP 4.0, task directives were extended to include dependency clauses.
Expressing the data dependencies of the BLAS operations directly in the \lstinline[style=inlinecpp]|depend| clauses allows the runtime to schedule tasks as soon as their inputs are ready, so global synchronization barriers can be removed entirely.
The resulting task graph for the $4 \times 4$ tiled right-looking Cholesky is shown on the right-hand side of Figure~\ref{fig:variants}, with arrows encoding the data dependencies between the BLAS operations.

Based on these strategies, we evaluate four OpenMP implementation variants:
\begin{itemize}
	\item \textit{Fork-join}: a naive implementation parallelizing the outer loops in lines 3 and 6 of Figure~\ref{fig:tiled-algo} with static scheduling for line 3 and dynamic scheduling for line 6.
	\item \textit{Fork-join collapsed}: an optimized implementation that collapses the two submatrix-update loops (lines 6--8 of Figure~\ref{fig:tiled-algo}) into a single parallel loop nest.
	\item \textit{Task synchronous}: a task-based implementation using OpenMP~3.0 with explicit \lstinline[style=inlinecpp]|taskwait| synchronization.
	\item \textit{Task asynchronous}: a task-based implementation using OpenMP~4.0 task dependencies.
\end{itemize}

Annotating tasks with priorities (OpenMP~4.5), which hint to the scheduler at critical-path tasks, did not measurably improve the \textit{Task asynchronous} variant, so we report results without them.

\paragraph{HPX variants.}

In contrast to OpenMP, HPX is designed from the outset around asynchronous tasking.
Data dependencies are expressed naturally by chaining future-based data structures.
Futures can be chained via the \lstinline[style=inlinecpp]|.then(...)| and \lstinline[style=inlinecpp]|hpx::async(...)| constructs, or via \lstinline[style=inlinecpp]|hpx::dataflow(...)|, which manages the dependencies implicitly.
The programmer can choose between two dependency styles: using lightweight \lstinline[style=inlinecpp]|hpx::future<void>| handles for dependency tracking only, or embedding the tile data itself inside the future wrapper.
The first style is low-overhead, but since the futures do not own the data, an incorrect dependency specification can easily lead to data races and silently corrupt the result.
The second style is safer and syntactically cleaner, as the full task futurization can be expressed directly, at the cost of a small indirection whenever the data is accessed through the wrapper.
An \lstinline[style=inlinecpp]|hpx::future<T>| is consumed only once, enabling ownership transfer via move semantics. Tiles read by multiple tasks instead use \lstinline[style=inlinecpp]|hpx::shared_future<T>|, where modifying the data incurs an additional copy.

The underlying BLAS libraries follow the cBLAS interface and thus require raw pointers into contiguous memory.
To avoid error-prone manual buffer management, we store each tile as a \lstinline[style=inlinecpp]|std::vector<T>|, so each HPX task future wraps a \lstinline[style=inlinecpp]|std::vector| whose \lstinline[style=inlinecpp]|.data()| yields the pointer cBLAS expects.

To match the OpenMP study, we evaluate four HPX implementation variants:
\begin{itemize}
	\item \textit{Task asynchronous}: a task-based implementation using HPX dataflow and BLAS future wrappers, with dependencies expressed directly through the future chain.
	\item \textit{Task synchronous (futures)}: a task-based implementation using HPX dataflow and BLAS future wrappers that additionally enforces the global synchronization barriers of synchronous OpenMP-3.0-style tasking.
	\item \textit{Fork-join}: a naive implementation that parallelizes only the outer submatrix-update loop using \lstinline[style=inlinecpp]|hpx::experimental::|\allowbreak\lstinline[style=inlinecpp]|for_loop| with dynamic scheduling.
	\item \textit{Fork-join collapsed}: an optimized implementation that parallelizes both submatrix-update loops with a nested pair of such loops, applying dynamic scheduling to the outer loop and the runtime default to the inner one.
\end{itemize}

\section{Results}\label{sec:results}

In this section, we present benchmark results for the eight different parallelization variants of the tiled Cholesky decomposition.
All experiments are run on a single compute node equipped with two AMD EPYC 7742 (Zen~2) processors in a dual-socket configuration, for a total of 128 physical cores.
Reported runtimes are medians over 20 independent runs with FP64 double precision, which is robust to one-sided OS and scheduler noise. The error bars in the figures span the minimum and maximum observed runtimes.
We use randomly generated symmetric positive-definite matrices with per-dimension sizes ranging from $2^{8}$ to $2^{16}$, and we vary the number of tiles per dimension from four to $1024$.

All benchmarks are executed with 128 OpenMP or HPX worker threads so that both runtimes have access to the full set of physical cores on the node.
Sweeping the number of tiles per dimension lets us identify the configuration that best balances two competing effects: for a small number of tiles, there is insufficient parallel work to keep all 128 cores busy, whereas too many tiles produce fine-grained tasks whose execution time is dominated by runtime overhead.
For OpenMP, we pin threads with \texttt{OMP\_PROC\_BIND=close} and \texttt{OMP\_PLACES=cores}, which yielded the best performance in our preliminary experiments.
HPX worker threads are pinned to one physical core each.
All BLAS calls within a task are executed in sequential mode, so the OpenMP/HPX runtime is the sole source of parallelism.

Section~\ref{sec:tile_scaling} compares tile-size scaling for fork-join and tasking, both within and across the two runtimes.
In Section~\ref{sec:problem_size}, we then fix four representative tile counts and study problem-size scaling to isolate the tasking overheads that dominate when there are too many small tasks.
Section~\ref{sec:compiler_comparison} closes with a comparison of the OpenMP libraries shipped with GCC and LLVM.
Detailed software versions and build configurations are listed in the \hyperref[app:materials]{Supplementary Materials}.
\subsection{Tile Size Scaling}\label{sec:tile_scaling}

\begin{figure}[t]
	\centering
	\begin{minipage}[t]{.47\textwidth}
		\centering
		\includegraphics[width=\linewidth]{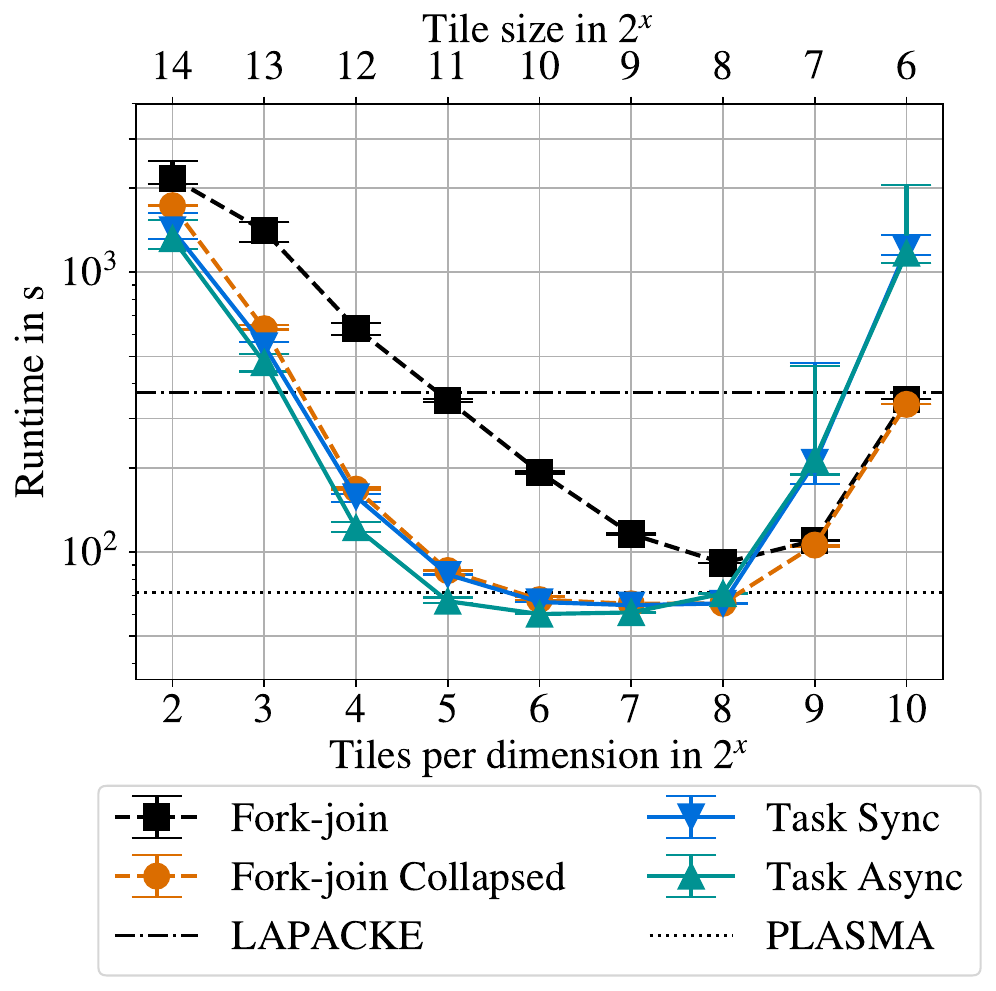}
		\caption{Tile-size scaling of the four OpenMP fork-join and tasking variants for a problem size of $2^{16}$ on 128 OpenMP threads, compiled with GCC.}
		\label{fig:tile_size_omp}
	\end{minipage}
	\hspace{0.5cm}
	\begin{minipage}[t]{.47\textwidth}
		\centering
		\includegraphics[width=\linewidth]{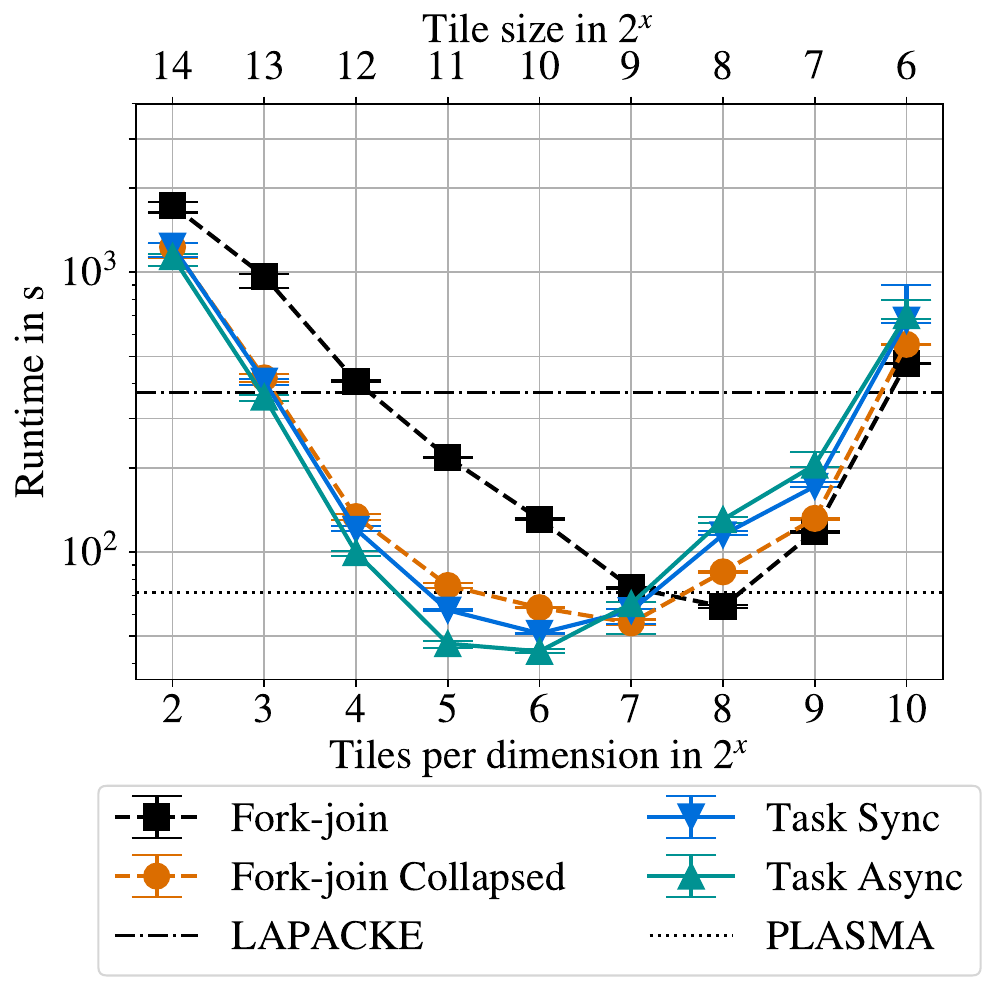}
		\caption{Tile-size scaling of the four HPX fork-join and tasking variants for a problem size of $2^{16}$ on 128 HPX OS-threads, compiled with GCC.}
		\label{fig:tile_size_hpx}
	\end{minipage}\hspace{.05\textwidth}
\end{figure}

Figure~\ref{fig:tile_size_omp} shows the tile-size scaling of the four OpenMP variants, compiled with GCC, for a problem size of $2^{16}$.
The two horizontal axes denote the number of tiles per matrix dimension and the resulting tile size.
Two dotted reference lines are added: LAPACKE backed by parallel OpenBLAS~\cite{Zhang2025_openblas}, representing a non-tiled call into a multi-threaded BLAS, and PLASMA~\cite{Dongarra2019_plasma_openmp}, an established tiled-Cholesky implementation that also relies on OpenMP tasking.
To avoid an integer overflow in PLASMA at its default tile size for this problem size, we run PLASMA on a slightly reduced problem of size $2^{16} - 2^{8}$, corresponding to 255 tiles per dimension instead of 256.
The curves exhibit the expected shape: there is a well-defined sweet spot at which there is enough parallel work to saturate all 128 cores while individual BLAS tasks remain large enough to amortize the task-management overhead.
The naive fork-join variant performs significantly worse than the other three up to 256 tiles per dimension because only the outer loop is parallelized, and there are too few iterations to feed all threads.
Collapsing the inner and outer submatrix-update loops yields a substantial improvement that effectively matches the performance of synchronous tasking.
Removing the synchronization barriers through asynchronous tasking provides a further performance gain.
The task-management overhead grows sharply once the tiles become too small.
At the optimal tile size, collapsing the inner loop yields a speedup of almost 30\% over the naive fork-join implementation and brings it on par with synchronous tasking.
Switching from synchronous to asynchronous OpenMP tasking then adds a further 7\% improvement.

Figure~\ref{fig:tile_size_hpx} shows the corresponding results for HPX. The overall trend mirrors the OpenMP case: naive fork-join is significantly slower until enough parallel work is available, tasking imposes more overhead than fork-join, and asynchronous tasking ultimately wins at the sweet spot.
Two differences are worth noting.
First, HPX's naive fork-join variant catches up with the collapsed variant more aggressively, closing the naive vs. collapsed gap.
Second, the overhead difference between fork-join and tasking is less pronounced than for OpenMP.
At the optimal tile size, synchronous HPX tasking is about 8\% faster than the collapsed HPX fork-join, and removing the redundant synchronization barriers yields an additional 14\% improvement for asynchronous tasking, which is the intended operation mode for HPX.

Comparing the absolute runtimes of the OpenMP and HPX variants at their respective best tile sizes, HPX is consistently faster across the board.
Specifically, the HPX variants of fork-join, fork-join collapsed, synchronous tasking, and asynchronous tasking are 30\%, 15\%, 21\%, and 26\% faster than their OpenMP counterparts, respectively.

The cross-runtime gap must be interpreted with care.
Our setup holds the algorithm, the BLAS layer, the compiler, the thread-pinning granularity, and the problem generator fixed, but OpenMP and HPX differ structurally in how they schedule work.
For the fork-join variants in particular, this scheduling difference plausibly accounts for a sizable fraction of the 30\% and 15\% gaps, in addition to any intrinsic differences in runtime overhead.
We therefore read the cross-runtime numbers as a comparison of the two stacks as they are typically deployed.

Both Cholesky-Bench stacks outperform the PLASMA reference at their respective best tile sizes.
At PLASMA's default tile size for this problem (a side length of $2^{8}$), our OpenMP variants match PLASMA's performance, indicating that the gap stems from PLASMA's tile size rather than a fundamental implementation advantage.

\subsection{Problem Size Scaling}\label{sec:problem_size}

\begin{figure}[t]
	\centering
	\includegraphics[width=0.937\textwidth]{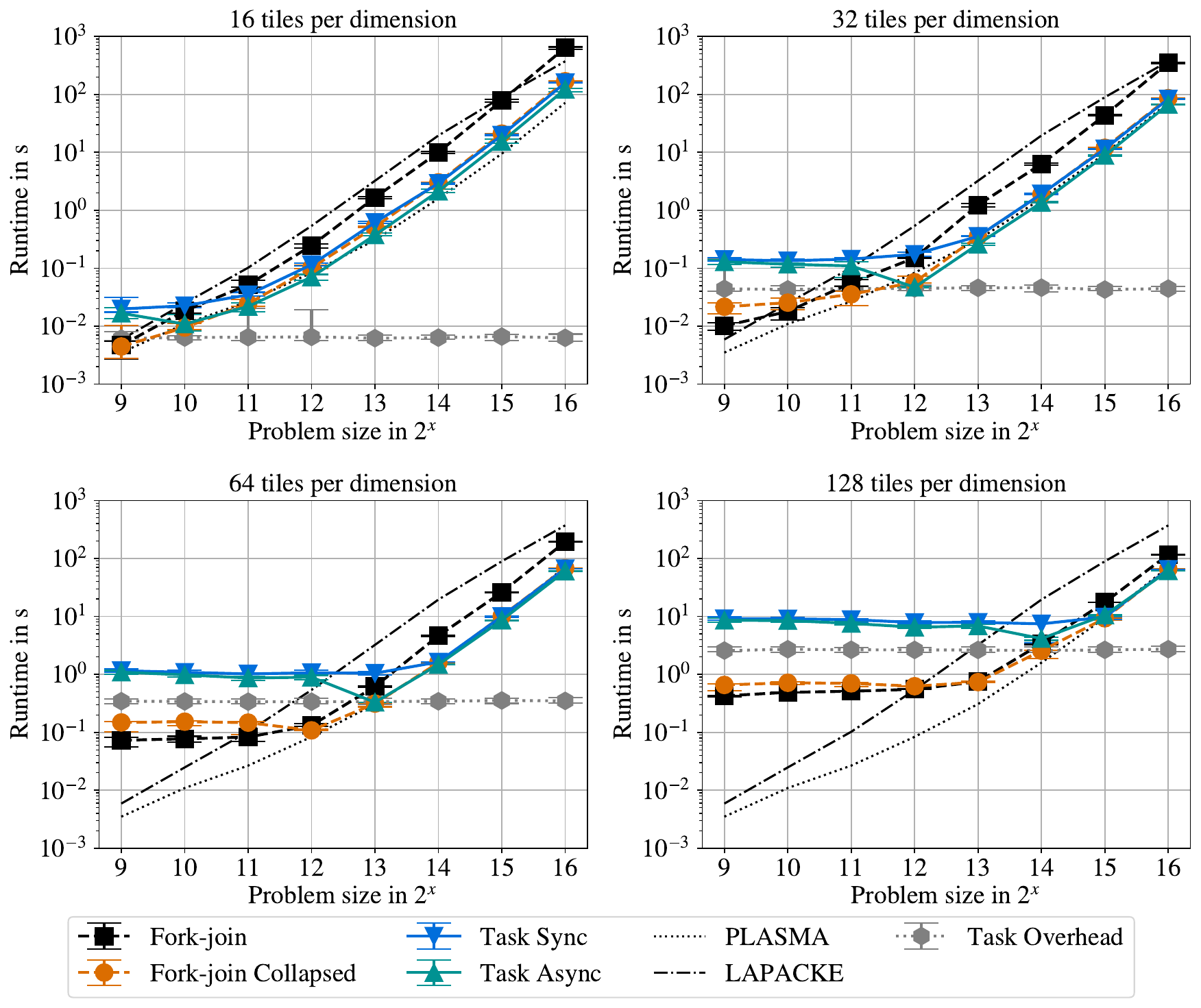}
	\caption{Problem-size scaling of the four OpenMP fork-join and tasking variants for 16, 32, 64, and 128 tiles per dimension on 128 OpenMP threads, compiled with GCC. The dashed curve labeled \textit{Task Overhead} shows the variant with all BLAS calls replaced by no-ops, isolating the pure task-management cost.}
	\label{fig:problem_size_omp}
\end{figure}

\begin{figure}[t]
	\centering
	\includegraphics[width=0.937\textwidth]{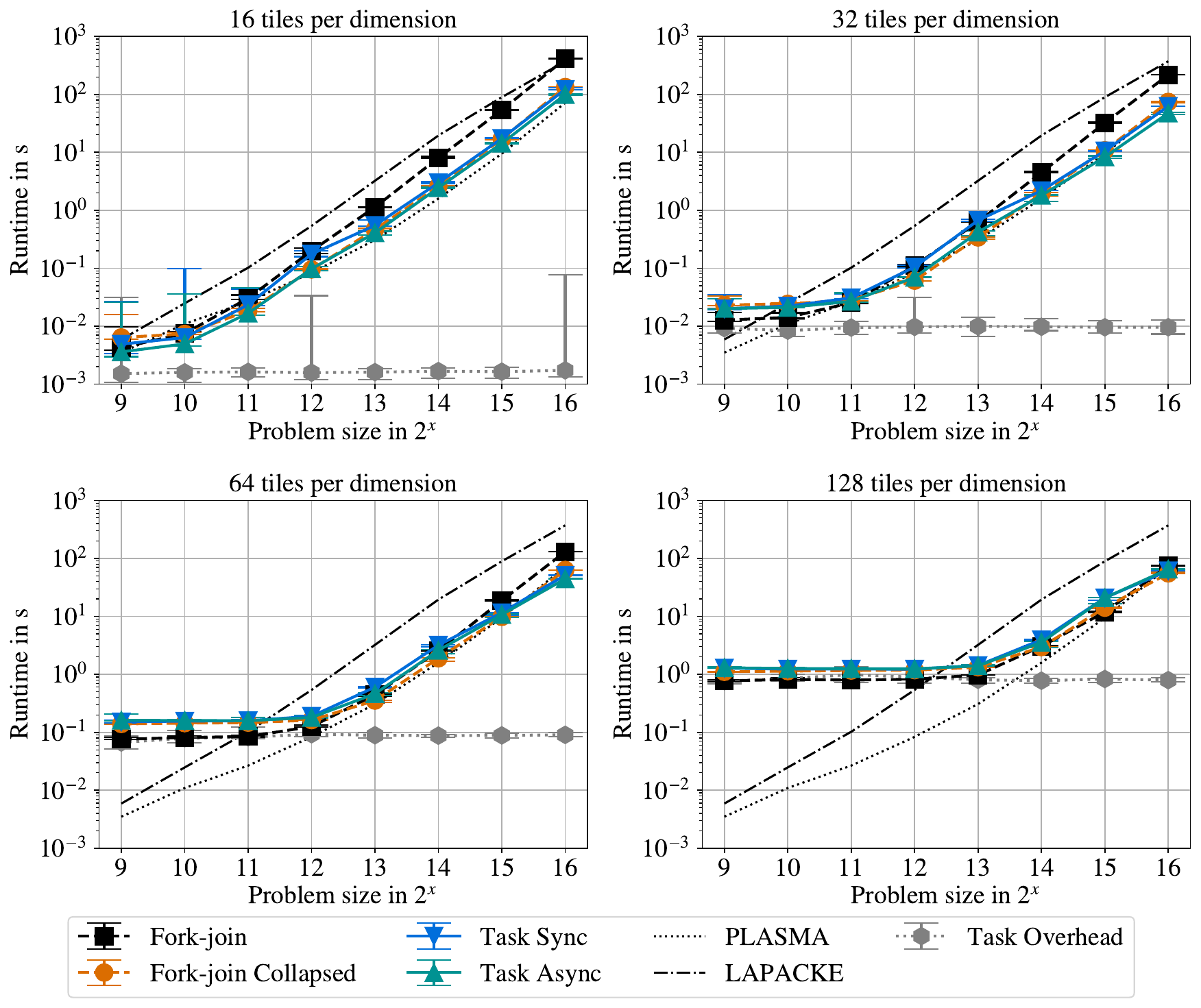}
	\caption{Problem-size scaling of the four HPX fork-join and tasking variants for 16, 32, 64, and 128 tiles per dimension on 128 HPX OS-threads. The dashed curve shows the \textit{Task asynchronous} variant with all BLAS calls replaced by no-ops, isolating the pure task-management cost.}
	\label{fig:problem_size_hpx}
\end{figure}

Figure~\ref{fig:problem_size_omp} reports problem-size scaling for a representative set of tile counts across the four OpenMP variants.
To isolate the runtime overhead of the \textit{Task asynchronous} variant from the BLAS work, we additionally plot a curve in which no-ops replace all BLAS calls.
This curve approximates the pure cost of task creation, dependency tracking, and scheduling.
Increasing the number of tiles improves performance at large problem sizes, but the same refinement also amplifies these runtime overheads for the task-based variants.
OpenMP tasking, in particular, incurs substantially higher overhead than fork-join, to the point that classical fork-join outperforms asynchronous tasking up to a certain problem size.
The plots also show that the optimal tile count depends on the problem size: for small problems, it is counterproductive to split the matrix so finely that all 128 cores are used.

Comparing these results with the corresponding HPX measurements in Figure~\ref{fig:problem_size_hpx} reveals a key difference: the lightweight tasks of HPX incur substantially lower overhead than OpenMP tasks across all tile counts.
As a consequence, for tile counts of 32, 64, and 128 per dimension, HPX asynchronous tasking dominates HPX fork-join across the entire problem-size range, with no crossover where fork-join would win.
Only at 16 tiles per dimension does the HPX overhead become small enough relative to the BLAS work that the two variants become indistinguishable.
To move beyond reporting runtimes, we use the no-op overhead curves to estimate the effective task overhead of each runtime.
For $n$ tiles per dimension, the right-looking factorization issues $n$ \textcolor{potrf}{\textbf{POTRF}}, $n(n-1)/2$ \textcolor{trsm}{\textbf{TRSM}} and \textcolor{syrk}{\textbf{SYRK}} each, and $n(n-1)(n-2)/6$ \textcolor{gemm}{\textbf{GEMM}} tasks.
Dividing the measured no-op runtime by this task count yields an effective task overhead that is nearly independent of the tile count, confirming that the runtime overhead grows linearly with the number of tasks.
For HPX this overhead is about $2\,\mu\mathrm{s}$ per task versus $7.6\,\mu\mathrm{s}$ for GCC OpenMP.
On this platform, this roughly $3.8\times$ smaller task overhead is consistent with the expected benefit of a runtime designed around asynchronous tasking, rather than the task model added to the fork-join-centric OpenMP standard. Establishing that this advantage holds across architectures and applications requires a broader study.

\subsection{Compiler Comparison}\label{sec:compiler_comparison}

\begin{figure}[t]
	\centering
	\includegraphics[width=0.75\textwidth]{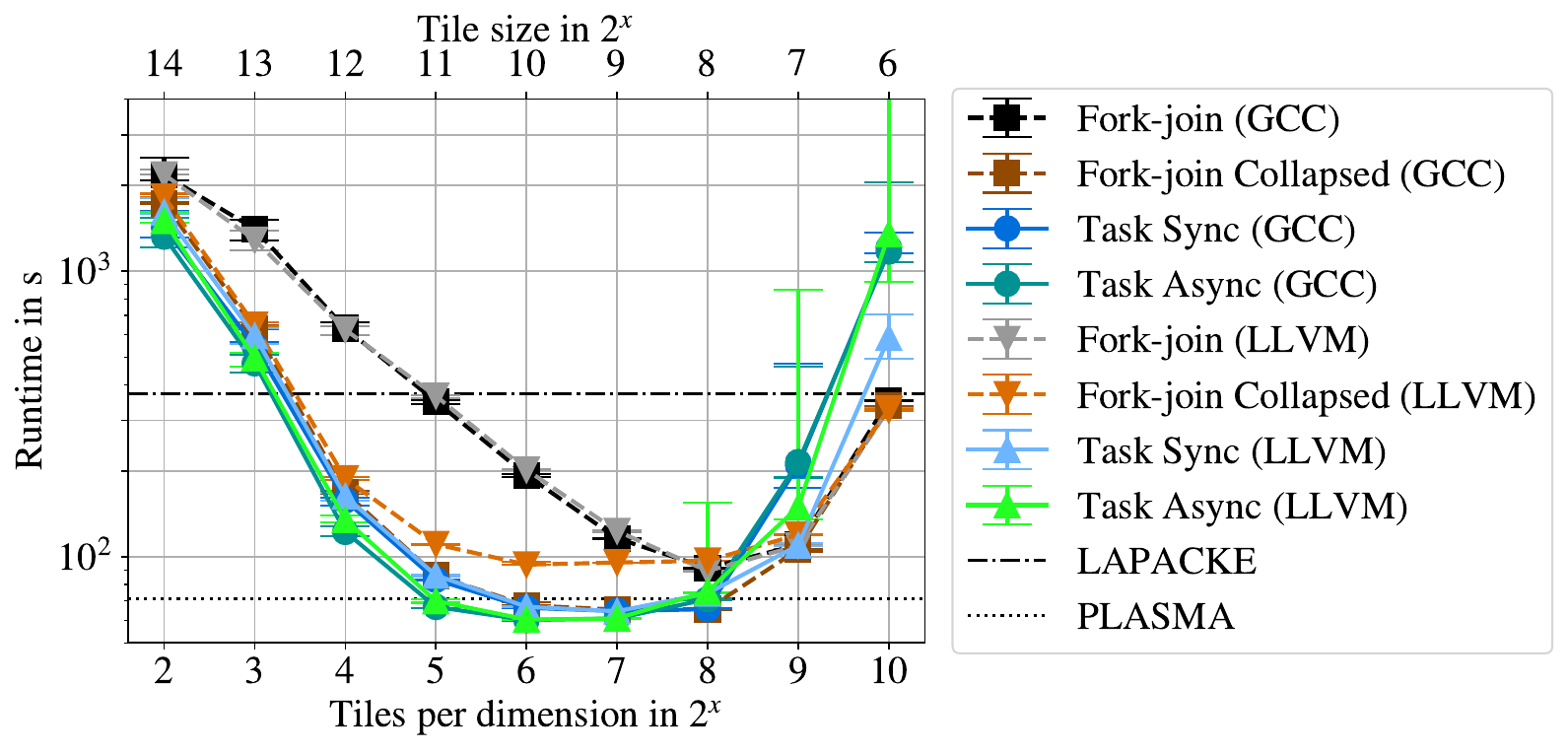}
	\caption{OpenMP tile-size scaling for a problem size of $2^{16}$, comparing the OpenMP runtimes of GCC 14.2.0 (OpenMP 4.5) and LLVM 22.1.2 (OpenMP 5.1) on the same source code.}
	\label{fig:compiler_comparison}
\end{figure}

Different compilers implement the OpenMP standard through different runtime libraries, and the choice of compiler can therefore substantially affect the measured performance.
To validate the OpenMP results obtained with GCC, we recompile the same code with a recent LLVM toolchain and rerun the tile-size sweep.
The results are shown in Figure~\ref{fig:compiler_comparison}.

Overall, both compilers achieve similar performance, but two observations stand out.
First, the collapsed fork-join variant scales worse with the number of tiles under LLVM than under GCC.
Second, the task-creation overhead for tasks without dependencies is noticeably lower for LLVM than for GCC.
At the optimal tile size of each variant, tasking performance and naive fork-join performance are essentially identical between the two compilers.
At the same time, GCC is 44\% faster on the collapsed fork-join variant.

The 44\% gap in the collapsed variant stems from a divergence between the two implementations of the OpenMP standard.
Because the fused trailing-update loop is non-rectangular, the OpenMP specification does not permit attaching an explicit \lstinline[style=inlinecpp]|schedule| clause to the surrounding \lstinline[style=inlinecpp]|collapse(2)| construct.
GCC enforces this restriction strictly: adding \lstinline[style=inlinecpp]|schedule(dynamic)| produces a compilation error.
LLVM accepts the same clause as a non-standard extension, and when dynamic scheduling is enabled, the LLVM gap to GCC closes to roughly the level seen on the naive fork-join variant.
The numbers reported in Figure~\ref{fig:compiler_comparison} therefore reflect the standard-conforming code path for both compilers.

\section{Conclusion and Outlook}\label{sec:conclusion}

In this work, we presented Cholesky-Bench and used it to compare fork-join and task-based parallelization variants of the tiled right-looking Cholesky decomposition using the OpenMP libraries shipped with GCC and LLVM, and the HPX runtime.
Tile-size and problem-size scaling benchmarks on a dual-socket 128-core AMD Zen~2 node reveal that the collapsed fork-join variant alleviates most of the load-imbalance bottleneck in the inhomogeneous Cholesky workload and becomes competitive with synchronous tasking.
Removing the remaining synchronization barriers through asynchronous tasking yields a further 7\% improvement for OpenMP and 14\% for HPX.
At their respective optimal tile sizes, all four HPX variants outperform their OpenMP counterparts.
Asynchronous HPX tasking in particular is more than 26\% faster than its OpenMP analog.
Beyond more effective scheduling, HPX also exhibits consistently lower task management overhead.

For small problem sizes, however, fork-join can even outperform asynchronous tasking.
The collapsed fork-join variant, in particular, remains a strong alternative that combines low overhead for small inputs with performance comparable to synchronous tasking for large inputs.
Nevertheless, when configured with a suitable tile size, asynchronous HPX tasking achieves the best performance across all problem sizes considered.

A limitation of our study is that all measurements were collected on a single dual-socket AMD Zen~2 node, so the precise speedups reported here are platform-specific.
Confirming our findings on further microarchitectures, such as Intel, ARM, and emerging RISC-V~\cite{Strack2026_gprat_riscv} platforms, is an important direction for future work.
Several other directions for future work are apparent.
We plan to extend Cholesky-Bench to include additional AMT runtimes, such as TTG or StarPU, to enable a fair comparison using the same software stack and to quantify their relative runtime overheads, in the spirit of Task Bench~\cite{Slaughter2020_taskbench}.
Additionally, we intend to add the left-looking and top-looking variants of the tiled Cholesky decomposition to assess whether the choice of algorithmic traversal affects our findings.
Finally, while shared-memory Cholesky captures only a small slice of the capabilities of modern AMTs, extending the study to a distributed setting and to other applications that combine irregular computation with communication is a natural next step.

\begin{credits}

	\phantomsection
	\subsection*{Supplementary Materials}\label{app:materials}

	Cholesky-Bench is publicly available as open-source software under the MIT License on \href{https://github.com/constracktor/Cholesky-Bench}{GitHub}\footnote{\url{https://github.com/constracktor/Cholesky-Bench}, accessed 2026-06-10.}.
	The repository includes compilation and \texttt{sbatch} scripts that can be extended to additional hardware.
	All experiments were conducted on the configuration summarized below; unspecified environment variables are left at their defaults.
	\begin{itemize}
		\item \textbf{Hardware:} dual-socket node with two AMD EPYC 7742 processors (Zen~2, 64~cores / 128~threads per socket; SMT enabled but worker threads pinned one-per-physical-core) and 2046~GiB DDR4 RAM.
		\item \textbf{Operating system:} Ubuntu 24.04.4 LTS, Linux kernel 6.8.0-106-generic.
		\item \textbf{Compilers:} GCC 14.2.0 (OpenMP 4.5) and LLVM/Clang 22.1.2 (OpenMP 5.1).
		\item \textbf{Build configuration:} All Cholesky-Bench code was compiled in release mode (\texttt{-O3 -DNDEBUG}). OpenBLAS, when installed via Spack, is automatically compiled for the target Zen~2 microarchitecture.
		\item \textbf{HPX:} version 1.11.0, built with jemalloc as the memory allocator.
		\item \textbf{BLAS / LAPACKE:} OpenBLAS~\cite{Zhang2025_openblas} 0.3.28, invoked through the cBLAS interface. Two builds are used: a sequential build inside Cholesky-Bench tasks (so that the parallel runtime is the sole source of parallelism), and an OpenMP-threaded 64-bit-integer build for the LAPACKE and PLASMA reference lines.
		\item \textbf{PLASMA:} version 24.8.7~\cite{Dongarra2019_plasma_openmp}, built against the OpenMP-threaded OpenBLAS 0.3.28 above.
	\end{itemize}

	\subsection*{AI Use Disclosure}

	Generative AI tools, including Grammarly~\cite{grammarly}, ChatGPT~\cite{chatgpt}, and Claude~\cite{claude}, were employed to enhance the clarity, grammar, and overall coherence of the manuscript. All technical content, data analyses, and research findings were conceived and developed independently by the authors. AI-assisted outputs were carefully reviewed, verified, and edited by the authors to ensure factual accuracy, interpretive rigor, and scholarly integrity. The final manuscript reflects the authors' original intellectual contributions and analytical work.

\end{credits}

\bibliographystyle{splncs04}
\bibliography{main}
\end{document}